# Running streams of a ferroelectric nematic liquid crystal on a lithium niobate surface

Luka Cmok[a], Virginie Coda[b], Nerea Sebastián[a], Alenka Mertelj[a], Marko Zgonik[a], Satoshi Aya[c], Mingjun Huang[c], Germano Montemezzani[b], and Irena Drevenšek-Olenik[a,d,*]

[a] J. Stefan Institute, Department of Complex Matter, Ljubljana, Slovenia [b]Université de Lorraine, Centrale Supélec, LMOPS, Metz, France [c]South China University of Technology, Guangzhou, China [d]University of Ljubljana, Faculty of Mathematics and Physics, Ljubljana, Slovenia

*irena.drevensek@ijs.si

# Running streams of a ferroelectric nematic liquid crystal on a lithium niobate surface

Sessile droplets of a ferroelectric nematic liquid crystalline material were exposed to surface electric fields produced by pyroelectric and photogalvanic (photovoltaic) effects in X-cut iron-doped lithium niobate crystals. The resulting dynamic processes were monitored by polarization optical (video)microscopy (POM). During heating/cooling cycles, at first, the droplets change their shape from spherical to extended ellipsoidal. Then they start to move rapidly along the surface electric field, i.e., along the crystal's polar axis (c-axis). During this motion, several droplets merge into running streams (tendrils) extending towards the edges of the top surface area. Finally, practically all liquid crystalline material is transported from the top surface to the side surfaces of the crystal. At stabilized temperature, laser illumination of the assembly causes dynamic processes that are localized to the illuminated area. Also, in this case, the LC droplets merge into several tendril-like formations that are preferentially oriented along the c-axis of the crystal. The pattern of tendrils fluctuates with time, but it persists as long as the illumination is present. In this case, the LC material is transported between the central and the edge region of the illuminated area.

**Introduction**

Contactless manipulation and transportation of liquid materials on solid surfaces represent an attractive challenge in the development of digital microfluidics, in which conventionally, microdroplets of a selected liquid are manipulated by the electric voltage generated via a complex array of electrodes attached to the supporting substrate [1-3]. If the electrodes are fabricated from stimuli-responsive materials, e.g., photoconductive electrodes, the droplets can be indirectly regulated by contactless stimuli, such as optical illumination [4]. However, direct methods for manipulation of droplets via contactless stimuli are quite rare [5-7]. In this work, we report on direct contactless manipulation of droplets of a ferroelectric nematic liquid crystal on the surface of a solid ferroelectric substrate and demonstrate that the interaction between a liquid and a solid ferroelectric medium leads to various intriguing new phenomena.

The substrates used in our work were an iron-doped lithium niobate crystals ($LiNbO_3$:Fe, LN:Fe). LN:Fe crystals are renowned for their strong pyroelectric as well as photogalvanic effects [8-10]. By heating/cooling the crystal, a balance between the ferroelectric polarization and the compensating surface charge is temporarily broken, so a temporary voltage appears between the opposite crystal Z-planes. Alternatively, by illumination with light in the green spectral region, a spatial redistribution of charge carriers inside the crystal occurs, which takes place predominantly along the polar axis (c-axis) of the crystal. In both cases, a static space-charge electric field in the range of 10-100 kV/cm is readily generated inside the crystal [11,12]. The pyroelectric field is typically present in the entire crystal, while the photogalvanic field (also termed the photovoltaic field) is localized to the illuminated area [13]. These fields extend into the surrounding space around the crystal in the form of an evanescent field [14]. Via electrophoretic and/or dielectrophoretic forces, the evanescent field can cause trapping and manipulation of small objects deposited on the crystal surface and induce their (re)distribution. As this redistribution follows the illumination pattern, the effect is known as photovoltaic tweezers (PVT) [15,16].

While thermo- and opto-electronic manipulation of nano- and micro-particles of condensed materials with LN:Fe substrates were extensively investigated, reports on experiments with liquid materials are relatively rare. They are mainly focused on the control of droplets of pure water or aqueous dispersions of different compounds [7, 17-

21], which are reviewed in [22]. Studies involving unconventional liquids, such as liquid crystals (LCs), are even more scarce. The research with LC materials can be divided into two categories: manipulation of bulk LC layers and manipulation of LC droplets. Research in the former category is aimed mainly at imprinting reconfigurable optical patterns into the LC medium [23-35], while investigations in the latter category deal mainly with rearrangement of the LC droplets in response to photovoltaic or pyroelectric fields [13, 36-38].

All above-mentioned research on LC materials combined with the LN:Fe substrates was performed with conventional nematic LC compounds. In the traditional nematic phase, rod-shaped molecules are orientationally aligned with respect to each other. Nevertheless, the phase is not polar, as the probabilities that molecular dipoles are pointing in each of the two opposite directions associated with the alignment axis are the same [39]. But, recently, nematic LC compounds exhibiting polar (ferroelectric) orientational order were discovered [40-43]. The ferroelectric nematic LC phase is a proper ferroelectric fluid that is extremely sensitive to external electric fields [44-46]. Therefore, it is naturally anticipated that it should strongly interact with the photovoltaic/pyroelectric fields of the LN:Fe crystals. In this work, we report on the response of a ferroelectric nematic LC material known as DIO [40, 46], to the evanescent electric field on the surface of an X-cut LN:Fe crystal. The field was generated by heating/cooling of the assembly or by its laser irradiation at a fixed temperature. For the latter, a green laser beam with a Gaussian intensity profile was used. The experiments were performed at different temperatures corresponding to the ferroelectric nematic ($N_F$), splay nematic ($N_s$), and the standard nematic (N) phases of the LC material.

**Experimental**

The congruent LN:Fe crystals with a 0.03 or 0.05 mol% dopant concentration were grown by using the Czochralski method [47]. The "as-grown" crystals were polarized, cut to the rectangular prisms with dimensions of 9 × 4 × 6 mm$^3$ (X × Y × Z) and optically polished. Before the experiments, the prism was heated to 200°C, washed in acetone, and dried by the flow of nitrogen.

The ferroelectric nematic liquid crystalline material (DIO [40]) used in the experiments is in the solid phase at room temperature. During heating, it exhibits the following phase sequence: Cr → 65 °C → N$_s$ → 84.7 °C → N → 174 °C Iso, where Cr denotes the crystal phase, N$_s$ the splay nematic phase (originally denoted as M$_2$ [40]), N the standard nematic phase, and Iso the isotropic phase. During cooling, additionally, a monotropic ferroelectric nematic phase N$_F$ (originally denoted as MP [40]) appears. The phase sequence during cooling is: Iso → 174 °C → N → 84.5 °C → N$_s$ → 68.8 °C → N$_F$ → ~45 °C → Cr. The N$_F$ phase exhibits a spontaneous dielectric polarization of around 5 μC/cm$^2$ and very large effective dielectric constants of $\varepsilon_\parallel \sim 10^4$ and $\varepsilon_\perp \sim 10^3$ measured in directions parallel and perpendicular to the nematic director, respectively [40].

For experiments associated with pyroelectric voltage, a LN:Fe crystal with its YZ plane oriented upwards (i.e. in the X-cut plane orientation) was placed onto a supporting glass slide and a "flake" of powdered DIO was placed onto its top surface. Then the assembly was put into the microscope heating stage (STC200, Instec Inc.) and slowly heated to the desired starting temperature. Subsequently the temperature was increased/decreased from the starting temperature to the chosen final temperature with a selected heating/cooling rate (typically in the range of 1-10 °C/min). After such a heating or cooling step, a lot of the LC material was usually transported from the top to the side and even to the bottom wall of the LN:Fe crystal, i.e., into the gap between the glass plate and the crystal. Therefore, in subsequent experiments, the LN:Fe crystal together with the adhered LC material was sometimes rotated upside down.

Before laser irradiation experiments, the assembly was left for a sufficiently long time at a fixed temperature to reach a stable state. *In-situ* optical irradiation was implemented by a CW laser beam operating at the wavelength of 532 nm that was directed onto the assembly via the arm for the episcopic illumination in the polarization optical microscope (Optiphot-2-Pol, Nikon). The spot size of the beam on the top surface was 0.8 mm, and its optical power was 40 mW. The beam was impinging in the center of the observation area.

The heating/cooling and the irradiation-induced modifications of the shape and positions of the LC droplets were monitored with polarization optical microscopy (POM) by using the diascopic illumination configuration. The unwanted reflected laser light was filtered away by an appropriate notch filter. In most of the experiments, the angle between the polarizer and the analyzer was set to be 70°. This allowed us to resolve more details than in the case of crossed polarizers.

In some experiments, the LN:Fe crystal was illuminated with laser light even before the LC material was placed on it. The pre-illumination was performed within a separate setup using a cylindrical focusing lens. This caused a modification of optical properties that were visible as a dark strip inside the crystal. However, we found that such pre-illumination did not have any significant effects on the LC droplets subsequently placed on the crystal. We assume that this happened because, during the time needed to move the crystal from the illumination setup to the optical microscope, photoinduced charges at the surface of the crystal were practically compensated by the atmospheric charges.

**Results and discussion**

*Effects of pyroelectric field*

The investigations started by placing a "flake" of DIO onto the top surface of the LN:Fe crystal at room temperature (23 °C). Then the assembly was heated to 110 °C at the rate of 5 °C/min. Figure 1 shows a set of subsequent images recorded during this process. We observed that already in the Cr phase (Figure 1(a)), the aggregates elongated along the c-axis of the LN:Fe crystal. As soon as the melting started, liquid streams formed that were running along the c-axis (Figures 1(b,c)). At the end of the heating process, only a few small droplets remained in the initial area (Figure 1(d)), while most of the LC material was moved on the side walls (Z-planes) and even to the bottom side of the crystal.

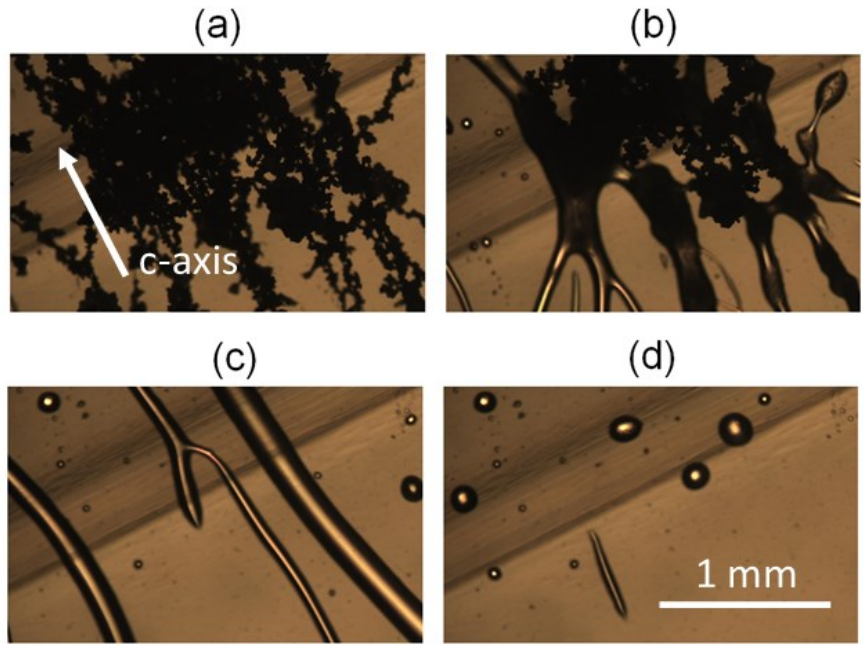

Figure 1. Surface transport of LC material (DIO) on X-cut LN:Fe crystal induced by heating from 23 °C to 110 °C by the rate of 5 °C/min. The corresponding pyroelectric field "pulls" the LC from the top to the side walls (Z-planes) of the crystal. (Dark strip-

shaped region on the substrate corresponds to the area that was pre-illuminated with the green laser beam)

In another experiment, the LC material was at first heated to 110 °C and then stabilized at this temperature. A region at which several larger LC droplets were present (together with several smaller ones) was located (Figure 2(a)). Then the temperature was reduced from 110 °C to 80 °C at the rate of 5 °C/min. At first, this caused elongation of the droplets in the direction of the c-axis (Figure 2(b)). Then the droplets merged together into the running streams by which the material was transported along the c-axis (Figure 2(c)). With increasing time, these streams became thinner and thinner (Figure 2(d)). At the end, again, only a few tiny droplets remained. In figure 2(c) one can notice that this image is significantly brighter than other images. This is because some LC material was transported into the gap between the bottom surface of the LN:Fe crystal and the supporting glass plate. Due to optical birefringence of this intruding LC layer, in polarization optical microscopy, more light is transmitted through the entire assembly.

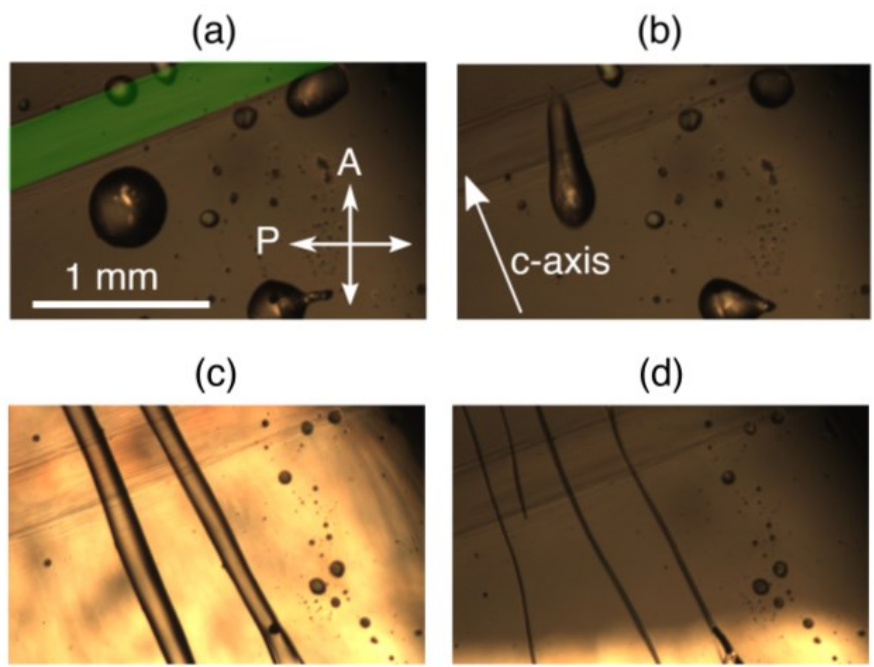

Figure 2. Droplets of LC material (DIO) on the LN:Fe crystal during cooling from 110 °C to 80 °C: (a) before cooling, (b) during cooling the droplets start to elongate, (c) then they abruptly transform into running streams, (d) the streams become thinner when the temperature starts to stabilize. In the top left corner (highlighted in green in Fig.1a) the region pre-illuminated by the 532-nm laser light can be seen. Image (c) is very bright in comparison to the others because a layer of LC material traveled into the gap between the bottom surface of the LN:Fe crystal and the underlying glass slide.

Interesting dynamic features can also be observed if one focuses on the LC material that moved into the gap between the bottom surface of LN:Fe crystal and the supporting glass surface. In particular, during cooling, when ferroelectric nematic phase $N_F$ is formed (T < 68 °C), several conical spikes appear at the edge region of the LC layer (Figure 3(a)). Such spikes also evolve from some of the LC droplets. During temperature variation, via those spikes, the droplets move around in a spider-like manner over distances that sometimes span the entire observation area (~ 2 × 2 mm$^2$). Two snapshots of such motion are shown in Figures 3(b) and (c). The motion of "spiders" exhibits no preferential direction.

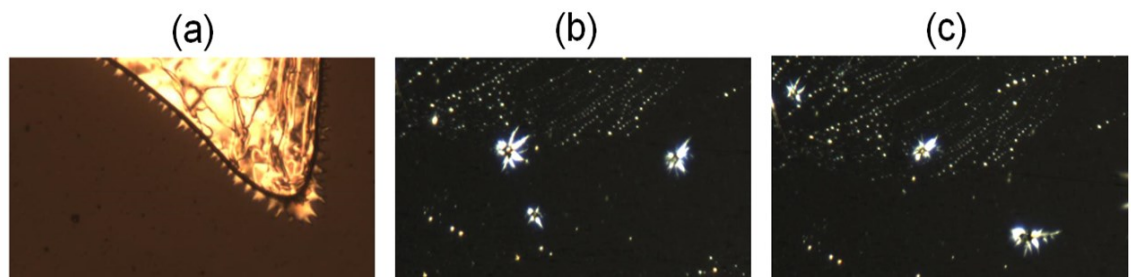

Figure 3. (a) Spiked conical formations observed at the edge of the LC layer trapped in-between the LN:Fe crystal and the glass plate during cooling into the $N_F$ phase. (b, c) The isolated spiky droplets move around in the image area.

The spiky formations observed in our experiments look very similar to the conical spikes that appear when (ferromagnetic) ferrofluid is exposed to the static magnetic field. They are associated with the so-called Rosensweig instability of ferrofluids [48,49]. In this work we explore ferroelectric ferrofluid exposed to the static electric field, so some similarities are not surprising. However, one should keep in mind that static electric fields are usually fast compensated by different kinds of charges present in the investigated system or its surrounding, while such compensation does not take place in magnetic systems.

Recently, Barboza et al. investigated large droplets (several mm in size) of another ferronematic LC compound (RM734, [41,42])) exposed to the pyroelectric field of a Z-cut LN:Fe crystal [50]. In the ferroelectric nematic phase, also they observed jetting effects generating tendril-like structures extending in the radial direction with respect to the droplets. This suggests that a tendency to form tendril-like arrangements represents a characteristic response of ferronematic LC droplets to a space-charge electric field of the LN:Fe crystals.

*Effects of photovoltaic field*

In contrast to heating/cooling, laser irradiation of LN:Fe crystals at a fixed temperature induces electric field only in the region of illumination. The illumination causes charge migration along the c-axis. In the open-circuit conditions, as used in our experiments, after some time, the current ends and stable spatial distribution of charges is generated. The resulting static photovoltaic electric field exhibits characteristics of the dipolar field with effective dipole moment oriented along the c-axis of the crystal [14]. Figure 4 shows black and white images of illumination-induced phenomena observed at 110 °C in the area initially involving only some relatively small LC droplets. The laser spot size is denoted by the green circle. During illumination, the LC material is "pulled" into the illuminated area and forms tendril-like structures that are preferentially oriented along the field lines of the photovoltaic field (Figure 4(b)). From time to time, some tendrils escape out of the illuminated area and form droplets at its edge, so fewer tendrils are observed inside the laser spot (Figure 4(c)). After some time, these droplets are pulled back and reformed into the tendrils. The described tendril-based pattern exists as long as the illumination is present. After switching off the laser beam, the tendrils transform back into droplets.

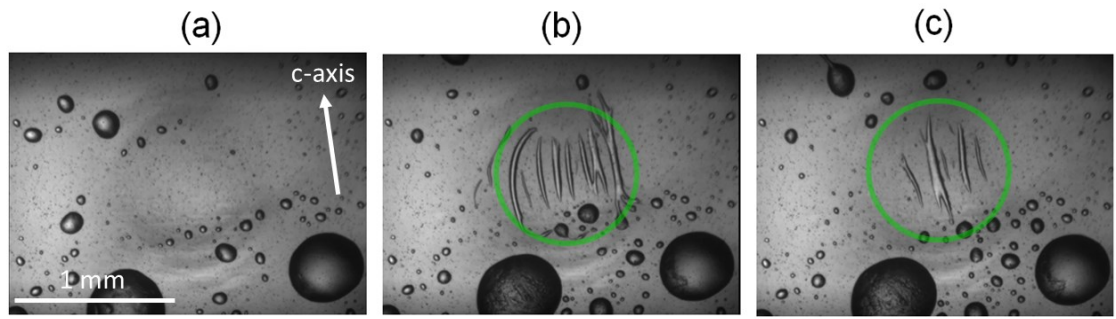

Figure 4. Droplets of LC material (DIO) on the X-cut LN:Fe crystal at 110 °C: (a) before illumination, (b, c) tendril-like structures formed during illumination. The green circles in (b) and (c) indicate the illuminated area.

In another experiment, the assembly was slowly cooled from 110 °C to 55 °C, i.e. into the ferroelectric nematic phase ($N_F$), and the experiment was repeated. In this process, the LC droplets in the illuminated area merged into laterally connected tendril-like formations that were as well oriented more or less parallel to the c-axis of the crystal (Figure 5(a)). After this, the laser beam was shortly switched off. During the "dark interval", the tendrils disconnected from each other (Figure 5(c)). Then the laser beam was switched on again, and the previous composition of laterally linked tendrils was reestablished (Figure 5(d)).

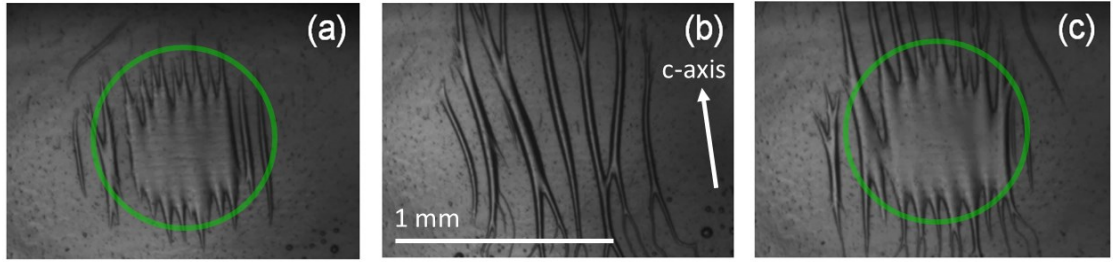

Figure 5. Illumination-induced tendrils of the LC material (DIO) formed on the surface of the X-cut LN:Fe crystal at 55°C: (a, c) during illumination, (b) when the illumination was shortly switched off. The green circles in (a) and (c) indicate the illuminated area.

In both above-described cases, the LC material forming droplets in size much smaller than the illuminated area is rearranged into tendril-like (filamented) structures that grow in a symmetric manner with respect to the c-axis of the crystal. This observation suggests that dielectrophoretic forces dominate over the electrophoretic forces [51]. Analogous phenomena were previously investigated with droplets of conventional dielectric liquids, such as water, exposed to strong electric fields [52-55]. The spikes, also known as Taylor cones, are attributed to the competition between the (di)electrophoretic forces and surface tension of the liquid [51]. With increasing electric field, the uncharged droplets stretch from spherical into ellipsoidal shape, and then Taylor cones (cusps) develop at their opposing ends. When the electric field reaches some critical value, deformed droplets become unstable and form the jet streams. One of the prominent differences between our work and previously reported experiments is that, in earlier works, to be able to observe the above-described phenomena on solid substrates, the substrates should be coated with a superhydrophobic coating to reduce the adhesion between the droplets and the substrate [7]. In our experiments, no coating was used at all, but the LC droplets were merging and moving on the untreated (bare) LN:Fe surface.

**Conclusions**

Our observations reveal, that when a set of several small droplets (diameter less than 1 mm) of the ferronematic LC material is placed on an X-cut LN:Fe crystal, the evanescent static electric field generated on its surface either via heating/cooling or via optical illumination, causes merging of the droplets into linear formations. In case of a pyroelectric field, which is present in the entire surface area, mass transport of the LC material along the direction of the field is generated, which causes the removal the LC material from the top surface of the crystal. In the case of a photovoltaic field, which is localized to the illumination area, the transport takes place only between the illuminated region and its neighborhood. The described phenomena are observed in the ferroelectric nematic phase, in the intermediate splay nematic phase, and in the standard nematic phase. Quite surprisingly, some effects were detected even in the isotropic phase. Further systematic experiments are needed to elucidate how the properties of different LC phases affect the details of the observed phenomena.

**Acknowledgements**

We thank R. A. Rupp for useful discussions and suggestions.

**Disclosure statement**

No potential conflict of interest was reported by the author(s).

**Funding**

This research was supported by the Slovenian Research Agency (ARRS) within the grant P1-0192 and bilateral exchange project BI-FR/19-20-PROTEUS-002.